\documentstyle[12pt]{article}

\catcode`\@=11
\long\def\@makefntext#1{
\protect\noindent \hbox to 3.2pt {\hskip-.9pt
$^{{\ninerm\@thefnmark}}$\hfil}#1\hfill}		

\def\@makefnmark{\hbox to 0pt{$^{\@thefnmark}$\hss}}  

\def\ps@myheadings{\let\@mkboth\@gobbletwo
\def\@oddhead{\hbox{}
\rightmark\hfil\ninerm\thepage}
\def\@oddfoot{}\def\@evenhead{\ninerm\thepage\hfil
\leftmark\hbox{}}\def\@evenfoot{}
\def\sectionmark##1{}\def\subsectionmark##1{}}
\renewcommand{\thefootnote}{\fnsymbol{footnote}}

\newcounter{sectionc}\newcounter{subsectionc}\newcounter{subsubsectionc}
\renewcommand{\section}[1] {\vspace*{0.6cm}\addtocounter{sectionc}{1}
\setcounter{subsectionc}{0}\setcounter{subsubsectionc}{0}\noindent
	{\normalsize\bf\thesectionc. #1}\par\vspace*{0.4cm}}
\renewcommand{\subsection}[1] {\vspace*{0.6cm}\addtocounter{subsectionc}{1}
	\setcounter{subsubsectionc}{0}\noindent
    {\normalsize\it\thesectionc.\thesubsectionc. #1}\par\vspace*{0.4cm}}
\renewcommand{\subsubsection}[1]
{\vspace*{0.6cm}\addtocounter{subsubsectionc}{1}
\noindent {\normalsize\rm\thesectionc.\thesubsectionc.\thesubsubsectionc.
    #1}\par\vspace*{0.4cm}}

\newcounter{appendixc}
\newcounter{subappendixc}[appendixc]
\newcounter{subsubappendixc}[subappendixc]

\renewcommand{\appendix}[1] {\vspace*{0.6cm}
        \refstepcounter{appendixc}
        \setcounter{figure}{0}
        \setcounter{table}{0}
        \setcounter{equation}{0}
        \renewcommand{\thefigure}{\Alph{appendixc}.\arabic{figure}}
        \renewcommand{\thetable}{\Alph{appendixc}.\arabic{table}}
        \renewcommand{\theappendixc}{\Alph{appendixc}}
        \renewcommand{\theequation}{\Alph{appendixc}.\arabic{equation}}
        \noindent{\bf Appendix \theappendixc #1}\par\vspace*{0.4cm}}

\renewenvironment{thebibliography}[1]
    {\begin{list}{\arabic{enumi}.}
    {\usecounter{enumi}\setlength{\parsep}{0pt}
\setlength{\leftmargin 1.25cm}{\rightmargin 0pt}
     \setlength{\itemsep}{0pt} \settowidth
    {\labelwidth}{#1.}\sloppy}}{\end{list}}

\newcounter{itemlistc}
\newcounter{romanlistc}
\newcounter{alphlistc}
\newcounter{arabiclistc}

\newcommand{\fcaption}[1]{
        \refstepcounter{figure}
        \setbox\@tempboxa = \hbox{\footnotesize Fig.~\thefigure. #1}
        \ifdim \wd\@tempboxa > 6in
           {\begin{center}
        \parbox{6in}{\footnotesize\baselineskip=15pt Fig.~\thefigure. #1}
            \end{center}}
        \else
             {\begin{center}
             {\footnotesize Fig.~\thefigure. #1}
              \end{center}}
        \fi}
\newcommand{\tcaption}[1]{
        \refstepcounter{table}
        \setbox\@tempboxa = \hbox{\footnotesize Table~\thetable. #1}
        \ifdim \wd\@tempboxa > 6in
           {\begin{center}
        \parbox{6in}{\footnotesize\baselineskip=15pt Table~\thetable. #1}
            \end{center}}
        \else
             {\begin{center}
             {\footnotesize Table~\thetable. #1}
              \end{center}}
        \fi}

 1 
 1  
 1  
 1  
 1  
 1  
 1  

\font\ninerm=cmr9

\begin{document}


February, 1995 \hfill  hep-ph/9503359
\begin{flushright}
VPI-IHEP-95-05\\
TUIMP-TH-95/65\\
MSUHEP-50315
\end{flushright}
\vskip 0.4in
\centerline{\normalsize\bf
EQUIVALENCE THEOREM AS A CRITERION FOR PROBING }
\baselineskip=17pt
\centerline{\normalsize\bf
THE ELECTROWEAK SYMMETRY BREAKING MECHANISM
\footnote{Talk presented by H.J.H.
at the {\it Beyond The Standard Model IV,}
December 13-18, 1994, Tahoe, California, USA.} }

\vspace*{0.9cm}
\centerline{\normalsize {\bf  Hong-Jian He} }
\baselineskip=17pt
\centerline{\normalsize\it
Department of Physics and Institute of High Energy Physics}
\baselineskip=14pt
\centerline{\normalsize\it
Virginia Polytechnic Institute and State University}
\baselineskip=14pt
\centerline{\normalsize\it Blacksburg, Virginia 24061-0435, USA}
\centerline{\footnotesize E-mail: hjhe@vtinte.phys.vt.edu}
\vspace*{0.4cm}
\vspace*{0.1cm}
\centerline{\normalsize\bf Yu-Ping Kuang}
\baselineskip=17pt
\centerline{\normalsize\it
CCAST ( World Laboratory ), P.O.Box 8730, Beijing 100080, China }
\baselineskip=14pt
\centerline{\normalsize\it
Institute of Modern Physics, Tsinghua University, Beijing 100084, China}
\centerline{\footnotesize E-mail: dmpkyp@tsinghua.edu.cn}
\vspace*{0.4cm}
\vspace*{0.1cm}
\centerline{\normalsize\bf C.--P. Yuan}
\baselineskip=17pt
\centerline{\normalsize\it
Department of Physics and Astronomy, Michigan State University }
\baselineskip=14pt
\centerline{\normalsize\it East Lansing, Michigan 48824, USA}
\centerline{\footnotesize E-mail: yuan@msupa.pa.msu.edu}

\baselineskip=18pt
\vspace{0.8cm}
\begin{abstract}

\noindent
Based upon our recent study on the Lorentz non-invariance ambiguity
in the longitudinal weak-boson scatterings  and the precise conditions for
the validity of the Equivalence Theorem (ET), we further examine the intrinsic
connection between the longitudinal weak-boson
scatterings and probing the electroweak symmetry breaking (EWSB)
mechanism. We reveal the profound physical content of
the ET as being able to discriminate processes which are insensitive to probing
the EWSB sector. With this physical content as a criterion, we analyze
the sensitivities to various effective operators for probing the mechanism of
the EWSB.

\end{abstract}

\setcounter{footnote}{00}
\renewcommand{\thefootnote}{\arabic{footnote}}
\renewcommand{\baselinestretch}{1.3}

\newpage
\noindent
\section{Introduction}

Despite the astonishing success of the Standard Model (SM) over the years,
its scalar part, the electroweak symmetry breaking (EWSB) sector,
remains as the greatest mystery.
Due to the screening theorem, the current low energy data,
allowing the SM Higgs boson mass to range from $60$\,GeV to about $1$\,TeV,
tell us little about the EWSB mechanism.
With the light Higgs particle(s) in the SM and SUSY-like theories remaining
un-detected, it is important to probe {\it all possible}
EWSB mechanisms: either weakly or strongly interacting.

While the transverse components $V_T^a$ of $~W^{\pm},~Z^0~$ are irrelevant to
the EWSB mechanism, the longitudinal weak-bosons
( $V_L^a=W^{\pm}_L,~Z^0_L~$), as the products of
the Higgs mechanism, are expected to be sensitive to
probing the EWSB sector. However, even for the strongly
coupled case, studying the $V_L$-scatterings does not guarantee
probing the EWSB sector in a sensitive and unambiguous way unless certain
general conditions are
satisfied to avoid the Lorentz non-invariance ambiguity of the
$V_L$-amplitudes \cite{et3}.
We note that the spin-$0$ Goldstone bosons (GB's) are
invariant under the proper Lorentz
transformations, in contrast, both $V_L$ and $V_T$
are Lorentz non-invariant
(LNI). After a Lorentz transformation, the $V_L$
 component can mix with  or even turn into a pure $V_T$.
Thus a conceptual and fundamental ambiguity arises: How can
we use the LNI $~V_L$-amplitudes to probe the
EWSB sector of which the physical mechanism should clearly be
independent of the choices of the Lorentz frames?
This motivated our precise formulation of the electroweak Equivalence
Theorem (ET) in Ref.[1].

The ET provides a quantitative
relation between the $V_L$-amplitude and the corresponding
GB-amplitude in the high energy region ($E\gg M_W$) \cite{et1,et2,et3};
the former is physically measurable while
the latter carries information of the EWSB sector.
Hence, as a bridge, the ET naturally connects
the $~V_L$-scattering experiments to probing
 the EWSB sector {\it in a precise way.}
As shown further below, the {\it difference}
between the $V_L$- and GB-amplitudes
is intrinsically related to the ambiguous LNI part of the $V_L$-scattering
which has the same origin as the $V_T$-amplitude,
and thus is insensitive to probing the EWSB sector.
When the LNI contributions can be safely ignored and the
Lorentz invariant (LI) scalar  GB-amplitude dominates
the experimentally measured $V_L$-amplitudes,
the physical $V_L$-scatterings can therefore
sensitively  and unambiguously probe the EWSB mechanism.
Furthermore,
in our precise formulation of the ET,
we show that the ET is not just a technical
tool in computing $V_L$-amplitudes via GB-amplitudes,
{\it as a criterion, it has an even more profound physical content for
being able to discriminate processes which are insensitive to probing
the EWSB sector}~\cite{et3}.

\section{The Precise Formulation of the ET for Probing the EWSB}

Starting from the Ward identity \cite{et1,et2}
{}~~$<0\,|F^{a_1}_0(k_1)\cdots F^{a_n}_0(k_n)\,
\Phi_{\alpha}|\,0>\,=\,0 ~$\footnote{
Here $~F_0^{a}~$ is the bare gauge fixing function and $\Phi_{\alpha}$
denotes other possible physical in/out states.}~
and making a rigorous LSZ reduction for the
external $F^a$-lines, we derived the following general identity for the
renormalized $S$-matrix elements:\footnote{
See the 2nd paper by H.-J. He, Y.-P. Kuang and X. Li in Ref.[3].}
$$
T[V^{a_1}_L,\cdots ,V^{a_n}_L;\Phi_{\alpha}]
= C\cdot T[-i\pi^{a_1},\cdots ,-i\pi^{a_n};\Phi_{\alpha}]+ B  ~~,
\eqno(1)                                               
$$
$$
\begin{array}{ll}
C & \equiv C^{a_1}_{mod}\cdots C^{a_n}_{mod} ~, \\
B   & \equiv\sum_{l=1}^n (~C^{a_{l+1}}_{mod}\cdots C^{a_n}_{mod}
T[v^{a_1},\cdots ,v^{a_l},-i\pi^{a_{l+1}},\cdots ,-i\pi^{a_n};\Phi_{\alpha}]
+ {\sf permutations~} )  ~,\\
v^a & \equiv v^{\mu}V^a_{\mu} ~,
{}~~~~~v^{\mu}~\equiv \epsilon^{\mu}_L-k^\mu /M_W = {\cal O}(M_W/E) ~,
\end{array}
\eqno(1a,b,c)                                            
$$
where $~\pi^a$'s  are GB fields, and the finite constant modification factor
$~C_{mod}^a~$ has been systematically studied in Ref.[3].
For clarity, let us assume that $~\Phi_\alpha~$ contains
either no field or some physical scalars
and/or photons. From (1), the LNI $~V_L$-amplitude can be decomposed
into two parts: the 1st part is
$~C\cdot T[-i\pi;\Phi_{\alpha}]~$ which is LI; the 2nd
part is the $~v_\mu$-suppressed $B$-term which is  LNI because of the external
{\hbox{spin-1}} $V_{\mu}$-field(s).  Such a decomposition
shows the {\it essential difference} between the $V_L$- and
the $V_T$-amplitudes since the former
contains a LI GB-amplitude which is the
intrinsic source causing a large $V_L$-amplitude in the case of
strongly coupled EWSB sector.
We note that only the LI part of the $V_L$-amplitude
is sensitive to probing the EWSB sector,
while its LNI part contains a
significant {\it Lorentz-frame-dependent} $~B$-term
and therefore is not sensitive to the EWSB mechanism.
Thus, for a sensitive and unambiguous probe of the EWSB,
we must find conditions for ignoring the $B$-term
such that the LI GB-amplitude
dominates the $V_L$-amplitude. {\it This physical content is essentially
independent of how to compute the $V_L$-amplitude.
It is the LI GB-amplitude that matters.}
It is clear that one can technically improve the prediction of the
$V_L$-amplitude from the RHS of (1) by including the complicated $B$-term ( or
part of $B$ ) \cite{gk}, but this is not an improvement of the equivalence for
$V_L$- and GB-amplitudes and thus irrelevant to the physical content of the ET
as a criterion for probing the EWSB mechanism.

{}From a detailed analysis \cite{et3} on the LNI $V_L$-amplitude,
we estimate the $B$-term as
$$
B \approx {\cal O}(\frac{M_W^2}{E_j^2})T[ -i\pi^{a_1},\cdots , -i\pi^{a_n};
  \Phi_{\alpha}] +
  {\cal O}(\frac{M_W}{E_j})T[ V_{T_j} ^{a_{r_1}}, -i\pi^{a_{r_2}},
                      \cdots , -i\pi^{a_{r_n}}; \Phi_{\alpha}]~~.
\eqno(2)                                                   
$$
We emphasize that the condition $~~ E_j \sim k_j  \gg  M_W , ~
(j=1,2,\cdots ,n) ~~$ {\it for each external longitudinal weak-boson} is
{\it necessary} for making the $B$-term ( and its Lorentz variation ) to be
much smaller than the GB-amplitude. This also precisely defines the {\it safe
Lorentz frames} in which the LNI
$B$-term can be ignored (cf. (3)). In conclusion, we give our general
and precise formulation of the ET as follows:
$$
T[V^{a_1}_L,\cdots ,V^{a_n}_L;\Phi_{\alpha}]
= C\cdot T[-i\pi^{a_1},\cdots ,-i\pi^{a_n};\Phi_{\alpha}]+
{\cal O}(M_W/E_j{\rm -suppressed} ),
\eqno(3)                                               
$$
$$
\begin{array}{l}
E_j \sim k_j  \gg  M_W , ~~~~~(~ j=1,2,\cdots ,n ~)~~;\\
B  \ll  C\cdot T[-i\pi^{a_1},\cdots ,-i\pi^{a_n};\Phi_{\alpha}] ~~,\\
\end{array}
\eqno(3a,b)                                   
$$
where {\it (3a,b) are the precise conditions for ignoring the LNI $B$-term
to validate the equivalence in Eq. (3).}
The amplitude $T$, to a finite order, can be written as
$~~T= \sum_{\ell=0}^N T_\ell = \sum_{\ell=0}^N \bar{T}_\ell \alpha^{\ell}~$
in the perturbative calculation.
Let $~~ T_0 > T_1,\cdots, T_N \geq T_{\min}~$, where
$~T_{min}= \{ T_0, \cdots , T_N \}_{\min}~$, then the condition (3b)
implies
$$
\begin{array}{l}
{\cal O}(\frac{M_W^2}{E_j^2}) \,T_0[ -i\pi^{a_1},\cdots , -i\pi^{a_n};
\Phi_{\alpha}] +
  {\cal O}(\frac{M_W}{E_j}) \,T_0[ V_{T_j}^{a_{r_1}}, -i\pi^{a_{r_2}},
                      \cdots , -i\pi^{a_{r_n}}; \Phi_{\alpha}] \\
 \ll  T_{min}[-i\pi^{a_1},\cdots ,-i\pi^{a_n};\Phi_{\alpha}]  ~~.
\end{array}
\eqno(4)                                                    
$$
We note that {\it the above formulation of the ET discriminates processes
which are insensitive to probing the EWSB sector when either
(3a) or (3b) fails.}
Furthermore, {\it as a physical criterion, the condition (4) determines
 whether or not the $~V_L$-scattering process
of interest is sensitive to probing
the EWSB sector to the desired precision in perturbative calculations.}

{}From (2) or the LHS of (4) and
the precise electroweak power counting rules \cite{hky}, we can
easily estimate the $B$-term to be
$$
B = {\cal O}(g^2)f_{\pi}^{4-n}~~
\eqno(5)                                     
$$
for theories with strongly coupled EWSB sector ( {\it i.e.},
the heavy Higgs SM or the chiral
Lagrangian formulated electroweak theories (CLEWT) ).
It is of the same order in magnitude as the
leading $V_T$-amplitude $T_0[V_T^{a_1},\cdots ,V_T^{a_n}]$.
Since both the $B$-term and the leading $V_T$-amplitude are of
order $g^2$~,
they are therefore insensitive to the EWSB sector
in accordance with the above general analysis.
If we want to probe the leading new physics contributions  in the CLEWT
at the $E^4$-level,
of $~{\cal O}(\frac{E^2}{f_{\pi}^2}\frac{E^2}{\Lambda^2}
f_{\pi}^{4-n})~$\footnote{In the CLEWT, $~~f_{\pi}=246$ GeV and the effective
cut-off $~\Lambda \simeq 4\pi f_{\pi}\simeq 3.1$ TeV.},~
then Eq. (5) and the criterion (4) yield
$~~~~
\frac{M_W^2}{E^2}  \ll  \frac{1}{4} \frac{E^2}{\Lambda^2} ~~,~~~{\rm or}~~~~
(0.70\,{\rm TeV}/E)^4  \ll  1 ~~. ~$
This shows that {\it in order to
sensitively probe the strongly coupled EWSB
sector, up to the order of $E^4$, we must measure the $V_L$ production
rates in the energy region above $1$TeV.}

\section{Sensitivities to the Effective Operators
         via Weak-Boson Scatterings }

Given the above conclusion, we examine the sensitivities to the
next-to-leading order effective operators \cite{eff},
 $~{\cal L}_{1,2}~$, $~{\cal L}_{9L,R}~$, $~{\cal L}_{10}~$ and
$~{\cal L}_{\Delta\rho}~$,
for probing the EWSB mechanism of the CLEWT via high energy
weak-boson scatterings. The coefficients of these operators
are of $~{\cal O}(1)\frac{f_{\pi}^2}{\Lambda^2}~$\footnote{
Due to the stringent experimental bound on $\Delta\rho$,
the coefficient of the dimension-$2$
operator $~{\cal L}_{\Delta\rho}~$ can also be at most
$~{\cal O}(1)\frac{f_{\pi}^2}{\Lambda^2}~$.} ~and {\it model-dependent}.

The condition (4) and Eq. (5) discriminate
which scattering process can
sensitively probe the EWSB sector at the
next-to-leading order in either hadron and electron collisions.
Define $R_L$ to be the ratio of $B$ ($\approx {\cal O}(g^2))$ to
${T_1[\pi\pi\rightarrow\pi\pi]}$, and
$R_T$ the ratio of $B$ to
${T_1[V_T\pi\rightarrow\pi\pi]}$.
At a given energy scale $E$, if $R_L$ (or $R_T$) is much less than one for
including the new physics contribution from the operator, say,
${\cal L}_{1}$, then we expect that this operator ({\it e.g.},
${\cal L}_{1}$) can be sensitively probed via the
scattering process $V_LV_L\rightarrow V_LV_L$
(or $V_TV_L\rightarrow V_LV_L$).
( We have assumed
the coefficient of each operator to be ${\cal O}(1)$,
after factorizing out the dimensional-counting
factor $\frac{f_{\pi}^2}{\Lambda^2}=\frac{1}{16 \pi^2}$. )
As summarized in Table 1, for $V_LV_L\rightarrow V_LV_L$ process, the
operators ${\cal L}_{1,2}$ can be sensitively probed
for $~E\geq 1$TeV~; while ${\cal L}_{9L,R}$ and ${\cal L}_{\Delta\rho}$
are insensitive even for $~E\approx \Lambda \sim 3$\,TeV~, where the
effective Lagrangian (${\cal L}_{eff}$) description becomes invalid.
${\cal L}_{10}$ has no contribution to this process at this order.
The operators  ${\cal L}_{9L}$ and ${\cal L}_{9R}$ are better probed
via $V_TV_L\rightarrow V_LV_L~(+ {\rm permutations})~$
than $V_LV_L\rightarrow V_LV_L$.
However,  $~{\cal L}_{10}~$ and $~{\cal L}_{\Delta\rho}~$
are totally insensitive via the $V_T$-processes.
A more complete discussion is given in Ref.[5].


\begin{table}[t]
\begin{center}
\tcaption{Sensitivities to the Next-to-Leading Order Effective Operators
          in $~{\cal L}_{eff}~$.}
\label{tab:exp}     

\vspace{0.5cm}

\small
\begin{tabular}{||c||c|c|c|c|c||}
\hline\hline
& & & & &  \\
Operators & ${\cal L}_1~,~{\cal L}_2$ & ${\cal L}_{9L}$  & ${\cal L}_{9R}$
& ${\cal L}_{10}$  & ${\cal L}_{\Delta\rho}$ \\
& & & & &  $(~ {\rm dim}=2~ )$ \\
\hline\hline
& & & & &\\
$T_1[\pi\pi\rightarrow\pi\pi ]$
& $\frac{E^2}{f_{\pi}^2}\frac{E^2}{\Lambda^2}$
& $g^2\frac{E^2}{\Lambda^2}$
&  $g^{\prime 2}\frac{E^2}{\Lambda^2}$
& /
& $\frac{E^2}{\Lambda^2}$  \\
& & & & &\\
\hline
& & & & &\\
$T_1[V_T\pi\rightarrow\pi\pi ] $
& $g\frac{E}{f_{\pi}}\frac{E^2}{\Lambda^2}$
& $g\frac{E}{f_{\pi}}\frac{E^2}{\Lambda^2}$
& $g^{\prime}\frac{E}{f_{\pi}}\frac{E^2}{\Lambda^2}$
& $gg^{\prime 2}\frac{f_{\pi}E}{\Lambda^2}$
& $g\frac{f_{\pi}E}{\Lambda^2}$   \\
& & & & & \\
\hline
& & & & & \\
$R_L \equiv \frac{B\approx {\cal O}(g^2)}{T_1[\pi\pi\rightarrow\pi\pi]}$
& $(\frac{0.70 TeV}{E})^4$
& $(\frac{3.1TeV}{E})^2 $
& $(\frac{5.7TeV}{E})^2$
& /
& $(\frac{2.0TeV}{E})^2 $  \\
& & & & &\\
\hline
& & & & & \\
$R_T \equiv \frac{B\approx {\cal O}(g^2)}{T_1[V_T\pi\rightarrow\pi\pi]}$
& $(\frac{1.15 TeV}{E})^3$
& $(\frac{1.15 TeV}{E})^3 $
& $(\frac{1.4  TeV}{E})^3$
& $\frac{200  TeV}{E}$
& $\frac{25  TeV}{E}$  \\
& & & & &\\
\hline\hline
\end{tabular}
\end{center}
\end{table}

\normalsize


\section{Acknowledgements}
H.J.H. thanks R. Casalbuoni, J. Gunion and T. Han for kind discussions
during the conference and the conference center for hospitality.
H.J.H. is supported in part by the U.S. DOE under grant
DEFG0592ER40709;
Y.P.K. is supported by the National NSF of China and
the FRF of Tsinghua University;
C.P.Y. is supported in part by NSF under grant PHY-9309902.

\end{document}